# High fidelity simulations of the multi-species Vlasov equation in the electro-static, collisional-less limit

R.-Paul Wilhelm*     Jan Eifert*     Manuel Torrilhon*

**Abstract**

The accurate prediction of occurrence and strength of kinetic instabilities in plasmas remains a significant challenge in nuclear fusion research. To accurately capture the plasma's dynamics one is required to solve the Vlasov equation for several species which, however, comes with a number of challenges as high dimensionality of the model as well as turbulence and development of fine but relevant structures in the distribution function. The predominantly employed *Particle-in-Cell* (PIC) method often lacks the accuracy to resolve the dynamics correctly, which can only be remedied by going to higher resolutions but at a prohibitorily high cost due to the high-dimensionality. Thus in this work we discuss the usage of the *Numerical Flow Iteration* (NuFI) as high fidelity approach, in contrast to e.g. PIC or grid-based approaches, to solve the multi-species Vlasov equation in modes leading to kinetic instabilities.

## 1 Introduction

The understanding of plasma dynamics in its kinetic regime is a key factor to better understand processes in many modern high-tech applications, in particular, to produce sustained fusion reactions in a controlled environment. Simulation of kinetic plasma physics comes with a number of challenges as turbulence and filamentation of the solution due to the high degree of non-linearity of the involved system, as well as the high dimensionality of the system due to modelling in the phase-space. While the former would require high resolutions to obtain accurate results, the later makes high resolution prohibetively expensive in terms of required resources and computation time due to the *curse of dimensionality* when working in the full six-dimensional phase-space.

Particle-In-Cell methods are predominantly used in Vlasov simulations but suffer from issues related to noise production in long-time simulations.[1],[2],[3],[4],[5],[6],[7],[8],[9] Therefore there has been active research on *noise-reduction techniques*[3],[4],[7],[10],[11] or combination to approaches using a mesh of the phase-space like the $\delta f$-approach.[12],[13],[14] While approaches using a full grid of the phase-space are known to be less prone to noise in general, they suffer from other artefacts like *numerical diffusion* which can be problematic in long-time simulations in the collisionless limit and leads to loss of a number of conservation properties.[15],[16],[17],[18],[19],[20],[21],[22] From a computational perspective grid-based approaches have to deal with meshing of the phase-space which can be challenging as soon as one has to deal with more complicated or changing geometries as well as adaptive grids to preserve accuracy. These challenges become even more problematic when one tries to simulate both electrons and ions kinetically (and potentially more than one species of ions). In this case one usually has to account for an even wider range of scales in the simulation as ions are several orders of magnitudes heavier than electrons. Additionally one often assumes different temperatures for electrons and ions, e.g. $T_e \gg T_i$ in case of ion acoustic shocks appearing in *laser driven shocks*.[1],[8],[9]

For these reasons there is an interest in looking into methods or models which reduce the dimensionality of the Vlasov system.[23],[24] The authors previously suggested a novel scheme, the Numerical Flow Iteration (NuFI), for solving the single-species Vlasov–Poisson system.[22],[25] This approach exploits the split structure of the velocity field in the Vlasov–Poisson equation to evaluate the distribution function $f$ via the characteristic map while only storing the electric potential $\varphi$ and thereby effectively reducing the dimension of the problem from 6 to 3 dimensions in terms of storage. On the one hand, this is essentially a shift from a memory- to a compute-bound approach using several orders of magnitude less memory for the same resolution, effectively enabling to use higher resolutions with the same hardware, however, at the cost of having quadratic computational complexity in the total number of time-steps instead of linear as in PIC. On the other hand, as one employs the characteristic map to reconstruct $f$ one avoids additional discretization errors in $f$ and preserves the solution structure. In particular, in previous works it has been shown that this not only leads to exact preservation of all $L^p$-norms, entropy, momentum and total energy (approximate but drift-free), but also leads to a better quality of results when comparing to other approaches with comparable resolution.[22],[25] Note that a similar approach, the *characteristic mapping method* (CMM), has been also investigated by Yin et.al. for the Vlasov–Poisson system and Euler system.[26],[27] They report similar results in terms of accuracy and structure-preservation though to a somewhat lesser extend as their characteristic map is directly stored leading to loss of incompressibility for the Vlasov–Poisson equation. Overall this makes NuFI a suitable candidate for high fidelity simulations in the simplified cases investigated so far, however, from a mathematical perspective it stands to reasons that these beneficial properties of NuFI in the simplified setting also carry over to more complicated settings. To this end we want to investigate how one can extend the idea of NuFI to a multi-species setting taking different types of boundary conditions on the system into account. The goal is to present the extensions and then show their applicability in the more general setting while comparing to a particle code to verify the results. We choose a particle-based approach to compare to NuFI due to its intrinsic adaption to both turbulence and changing velocity support, which can occur in the multi-species Vlasov setting.

---
*Applied and Computational Mathematics, RWTH Aachen University, Schinkelstraße 2, 52062 Aachen, Germany.
E-Mail: wilhelm@acom.rwth-aachen.de

In Section 2 we first present the Vlasov–Poisson system as well as NuFI in the periodic, single-species setting, before we move on to discussing extension towards coupling several particle species and handling of other boundary conditions. In Section 3 we then first show simulation results of NuFI for ion-acoustic waves in a periodic box, where we also compare the solution quality to results produced by PIC. Afterwards we also show results obtained by NuFI with Dirichlet boundary conditions before we summarize and discuss our results in Section 4.

## 2 Methods

In this section we present NuFI and the extension done towards efficiently handling multi-species systems. To keep the presentation comprehensive we restrict ourselves to $d = 1$ in this section, however, all of the following can be applied analogously to $d = 2, 3$ and the algorithms were developed keeping performance aspects in higher dimensions in mind.

### 2.1 Vlasov–Poisson system

In the Vlasov–Poisson system we represent each particle-species $\alpha$ through a corresponding distribution function $f^\alpha : \Omega \times \mathbb{R}^d \to \mathbb{R}_+$ with $\Omega \subset \mathbb{R}^d$, following

$$\partial_t f^\alpha + v \cdot \partial_x f^\alpha + \frac{q_\alpha}{m_\alpha} E \cdot \partial_v f^\alpha = 0, \tag{1}$$

where $q_\alpha$ and $m_\alpha$ are the corresponding charges and masses of the respective particle species.[28] The Vlasov equations for the different particle species are then coupled through a Poisson equation

$$E = -\nabla_x \varphi, \tag{2}$$

$$-\Delta_x \varphi = \rho, \tag{3}$$

$$\rho(t, x) = \sum_\alpha \int_{\mathbb{R}^d} f^\alpha(t, x, v) \mathrm{d}v. \tag{4}$$

While both the Vlasov–Poisson and Poisson equations are linear by themselves, the multiplication of $E$ and $\partial_v f^\alpha$ leads to non-linearity in the coupled system.

Note that, when we consider the periodic case with a single species, commonly electrons, we assume that the ions form a uniform background and therefore Equation (4) reduces to

$$\rho(t,x) = 1 - \int_{\mathbb{R}^d} f(t,x,v) \mathrm{d}v, \tag{5}$$

where we drop the $\alpha$ for brevity of notation. Furthermore, in the following we will assume that the involved quantities such as $f^\alpha$ and $E$ are normalized and dimensionless.[19]

## 2.2 Introduction to NuFI

The idea behind the *Numerical Flow Iteration* is to approximate the flow-map of the Vlasov–Poisson equation instead of directly solving for the distribution function $f$. This can be done using the Lagrangian perspective on the Vlasov–Poisson equation, i.e., the solution of (1) can be written as

$$f(t,x,v) = f_0(\Phi_t^0(x,v)), \tag{6}$$

where $s \mapsto \Phi_t^s(x,v) = (\hat{x}(s), \hat{v}(s))$ is the solution to

$$\begin{aligned} \frac{\mathrm{d}}{\mathrm{d}s}\hat{x}(s) &= \hat{v}(s), & \hat{x}(t) &= x, \\ \frac{\mathrm{d}}{\mathrm{d}s}\hat{v}(s) &= -E(s, \hat{x}(s)), & \hat{v}(t) &= v. \end{aligned} \tag{7}$$

We call $s \mapsto \Phi_t^s$ the *backward flow*. As the Vlasov–Poisson system is non-linear due to the coupling to Poisson's equation, one does not know $E$ a priori, however, it can be proven that the Vlasov–Poisson system indeed has a unique solution pair $(f, E)$ for sufficiently well-behaved initial data, which takes the above form.[29],[30] The inverse $s \mapsto \Phi_s^t = (\Phi_t^s)^{-1}$ of the backwards flow will be called the *forward flow*.

To solve (7) numerically we can employ the fact that the the velocity field $(v, -E(t,x))$ of the Vlasov–Poisson equation can be split into two components of which only $E$ depends on time. Thus to reconstruct the backwards flow it is actually sufficient to only know the electric field for the previous time-steps. To be precise, assuming for a moment that the electric field $E(t_i, \cdot)$ is known for $0 = t_0 < ... < t_i = i \cdot \Delta t < ... < t_n$ with $i < n \in \mathbb{N}$, then we can use the Störmer–Verlet time-integration scheme to solve (7) numerically. Starting from $\hat{x}_n^h = x$ and $\hat{v}_n^h = v$ compute

$$\begin{aligned} \hat{v}_{i-\frac{1}{2}}^h &= \hat{v}_i^h + \frac{\Delta t}{2} E(t_i, \hat{x}_i^h), \\ \hat{x}_{i-1}^h &= \hat{x}_i^h - \hat{v}_{i-\frac{1}{2}}^h, \\ \hat{v}_{i-1}^h &= \hat{v}_{i-\frac{1}{2}}^h + \frac{\Delta t}{2} E(t_{i-1}, \hat{x}_{i-1}^h) \end{aligned} \tag{8}$$

for $i = 0, ..., n$. It follows that $f(t, x, v) \approx f_0(\hat{x}_0^h, \hat{v}_0^h)$ up to an error of $\mathcal{O}(\Delta t^2)$.

As the electric field is not known a priori, to go from $t_{n-1}$ to $t_n$ one has to solve for the next electric field $E(t_n, \cdot)$, given the knowledge of the previous electric fields $E(t_i, \cdot)$ with $i \leq n-1$. To this end one has to first evaluate $\rho$, which involves computing the integral (5). In (8) one uses the yet unknown $E(t_n, \cdot)$, however, this can be circumvented by the following transformation of variable

$$\begin{aligned} \int_{\mathbb{R}} f_0(\Phi_0^{t_n}(x,v)) \mathrm{d}v &= \int_{\mathbb{R}} f_0 \left( \Phi_0^{t_{n-1}} \circ \Phi_{\text{half}}^1 \circ \Phi_{\text{half}}^2 \left( x, v + \frac{\Delta t}{2} E(t_n, x) \right) \right) \mathrm{d}v \\ &= \int_{\mathbb{R}} f_0 \left( \Phi_0^{t_{n-1}} \circ \Phi_{\text{half}}^1 \circ \Phi_{\text{half}}^2 (x, v) \right) \mathrm{d}v, \end{aligned} \tag{9}$$

where $\Phi_{\text{half}}^1$ is the step compute $\hat{v}_{n-1}^h$ and $\Phi_{\text{half}}^2$ the step to compute $\hat{x}_{n-1}^h$ from Equation (8). In the last step we used that the term $\frac{\Delta t}{2} E(t_n, x)$ is independent of $v$ thus $v \mapsto v + \frac{\Delta t}{2} E(t_n, x)$ has functional determinant 1 and we integrate over the whole space $\mathbb{R}$.

To solve the (3) for the electric potential, we first sample $\rho$ on an equi-distant grid in $x$ and then solve the Poisson's equation using e.g. *Fast Fourier Transformation* (FFT) in the periodic case or *Finite*

*Differences* (FD) in the Dirichlet case. The resulting $\varphi$ is then stored in a B-Spline basis, see Section 2.3. Note that while directly using the Fourier basis to evaluate $\varphi$ would have been possible, the involved exponential or trigonometric functions are more expensive to evaluate on a computer, thus in terms of speed of computations B-Splines are preferable over the Fourier basis.

For handling a multi-species Vlasov–Poisson system to compute $\rho$ one has to integrate over all species, i.e., one sums over the particle-species $\alpha$

$$\rho(t,x) = \sum_\alpha \int_{\mathbb{R}^d} f_\alpha(t,x,v) \mathrm{d}v. \tag{10}$$

Note that NuFI has an advantage over other classical approaches in this case: when going from one particle-species to two particle-species the computational and memory complexity essentially doubles for approaches directly storing $f$, while for NuFI only the computational complexity doubles but the memory complexity remains the same.

From a mathematical perspective NuFI entails several benefits in terms of structure-preservation over classical approaches.[22] Due to the use of a symplectic integrator in the backwards-in-time procedure (8) NuFI exactly preserves the *incompressibility* of the Vlasov–Poisson system, i.e., there is no numerical diffusion present and as a consequence all $L^p$-norms, the entropy, momentum are preserved exactly as well as total energy up to time-integration error. Additionally one is not restricted to an initially chosen resolution for $f$: as $f$ is evaluated through (6) on-the-fly one in principle can zoom into $f$ to an arbitrary high precision. Thus the integration procedure for $\rho$ can also be designed to be adaptive as well as account for changes in the velocity support, see also Section 2.5 and Section 2.6.

## 2.3 B-Splines for storage of potentials

To correctly model $\rho$ and $E$ using the previously computed electric potential $\varphi$, a high degree of smoothness is required for $\varphi$. This, together with the computational benefits of employing basis functions with local support, motivates the usage of B-Spline basis functions. For a periodic, one-dimensional domain $[0,L]$ fix a grid $s_{-k} < \ldots < s_0 = 0 < \ldots < s_{l+1} = L < \ldots < s_{l+k+1}$ with $k \in \mathbb{N}$. The B-Splines $N_{j,k}$ of order $k$ are recursively defined by

$$N_{j,1}(x) := \chi_{[t_j, t_{j+1})}(x), \quad N_{j,k}(x) := \frac{x-t_j}{t_{j+k-1}-t_j}N_{j,k-1}(x) + \frac{t_{j+k}-x}{t_{j+k}-t_{j+1}}N_{j+1,k-1}(x). \tag{11}$$

The interpolant of $\varphi$ is defined as

$$S(x) := \sum_{j=1}^k c_j N_{j,k}(x), \quad x \in [0,L], \tag{12}$$

with coefficients $c_j$ sought to satisfy the interpolation condition $S(x_j) = \varphi(x_j)$ inside the domain $[0,L]$.[31] Due to the basis functions only being defined locally, the resulting system of equations is of sparse nature, and thus fast iterative solver, such as LSMR, can be applied to efficiently compute the desired coefficients $c_j$.[32] An additional benefit is that once the coefficients are computed, the De-Boor formula allows for fast evaluation of $S(x)$ which does not require the evaluation for each basis function.[33] Furthermore, the derivatives of $S(x)$ which are required for $\rho$ and $E$ can be efficiently obtained as these derivatives are B-spline basis functions themselves and can be obtained in the same manner.

To impose boundary conditions, additional adjustments have to be made, depending on the type of condition. For higher order splines, special treatment is required for the last basis splines as their support leaves the domain $[0,L]$. In the case of periodic boundary conditions, this problem is solved by assigning the required $\varphi$ values outside of the domain via $\varphi(x_i) = \varphi(x_{i \bmod n})$, $i = n+1, .., n+k-1$, i.e., identifying nodes outside the domain with their periodic counterparts inside the domain.

To accommodate Dirichlet boundary conditions, the desired Dirichlet boundary conditions are imposed on $S(s_{-k}) = \ldots = S(s_0) = \varphi_{left}$ and $S(s_{l+1}) = \ldots = S(s_{l+k+1}) = \varphi_{right}$, as this removes any remaining degrees of freedom and forces $S(x)$ to interpolate $\varphi$ exactly for both the left and the right boundary while still keeping continuity. Analogously, the same ideas can be applied for periodic and Dirichlet boundary conditions in higher dimensions.

## 2.4 Handling of Dirichlet boundaries for $f$

Essentially the handling of boundary conditions with NuFI works analogously to particle methods like PIC: In the Lagrangian framework we consider the characteristics and have to check when they hit a boundary in $x$. In case of PIC one then would either remove the particle from the simulation if one has outflow conditions at that boundary or let it re-enter the domain (with a potentially modified weight) for periodic or in-flow conditions.[34] For NuFI, in case of Dirichlet boundaries we prescribe values at the boundaries, i.e., $f_{\partial\Omega}(t,x,v)$ for $x \in \partial\Omega$. Thus when a characteristic reaches $\partial\Omega$ during the backwards iteration (8), the scheme returns the respective value of $f_{\partial\Omega}$.

## 2.5 Adaptive integration for $\rho$

As mentioned in Section 2.2 there is no technical restriction on where one evaluates $f$ and how often one evaluates $f$, therefore one also is not really restricted in the choice of integration algorithm for $\rho$. Note, however, that in our experience the filamentation of $f$ for long-time simulations leads to issues with high-order integration methods due to the strong influence of higher derivatives on the error constant leading to large error constants. For this reason and for the sake of parallelization we suggested using the (equi-distant) mid-point integration rule for the more simple case of the single-species Vlasov–Poisson system in our previous work.[22]

When one is interested in long-time simulations, in particular for multi-species Vlasov systems, inhomogeneity of the distribution function $f$ becomes a considerable problem if not resolved correctly in the computation of $\rho$. To avoid having to run simulations with too high resolution in the beginning and insufficient resolution in later stages, one can use adaptive integration in $v$ to compute $\rho$. Keep in mind however, that we still want to avoid high-order quadrature rules and if possible re-use previous evaluations of $f$ as evaluating $f$ is the most expensive operation in NuFI. To this end we suggest using a combination of the trapezoidal and Simpson quadrature-rules. For given $t$ and $x$ fix $v_{\min}$ and $v_{\max}$, set $v_l = v_{\min}$, $v_r = v_{\min}$, compute $f_l = f(t, x, v_{\min})$ and $f_r = f(t, x, v_{\max})$. Then $\rho(t, x) = \text{INTEGRATE}(t, x, v_{\min}, v_{\max}, f_l, f_r, 1)$ with the function Integrate from Algorithm 1.

---

**Algorithm 1** Adaptive $\rho$ integration

   **function** INTEGRATE($t, x, v_l, v_r, f_l, f_r, i$)
        Compute $dv = v_r - v_l$.
        Compute $v_m = 0.5 \cdot (v_r + v_r)$.
        Compute $f_m = f(t, x, v_m)$.
        Compute $Q_T = 0.5 \cdot dv \cdot (f_l + f_r)$.
        Compute $Q_S = \frac{dv}{6} \cdot (f_l + 4 \cdot f_m + f_r)$.
        **if** $Q_S < \epsilon_1$ **then**
            **return** $Q_S$.
        **end if**
        **if** $|Q_S - Q_T|/Q_s < \epsilon_2$ or $i \geq \theta$ **then**
            **return** $Q_S$.
        **else**
            **return** INTEGRATE($t, x, v_l, v_m, f_l, f_m, i+1$) + INTEGRATE($t, x, v_m, v_r, f_m, f_r, i+1$).
        **end if**
   **end function**

---

We introduced the tolerance parameters $\epsilon_1, \epsilon_2 > 0$ and $\theta \in \mathbb{N}$:

- $\epsilon_1$ is used as tolerance to test for $Q_S \approx 0$ to not divide by something close to 0. Note that $f \geq 0$, thus $Q_S \geq 0$,

- $\epsilon_2$ is the precision-tolerance for integration,

- $\theta$ is the maximum depths of refinement allowed.

**Remark 2.1.** *Starting the adaptive integration from $v_l = v_{\min}$ and $v_r = v_{\max}$ might be a poor choice in practice as one runs the risk of missing fine structures due to a too coarse start grid. It is better to* start from a coarse initial grid $v_1 = v_{\min} < ... < v_m = v_{\max}$ with $m > 1$ and start Algorithm 1 in each sub-interval $[v_i, v_{i+1}]$ independently.

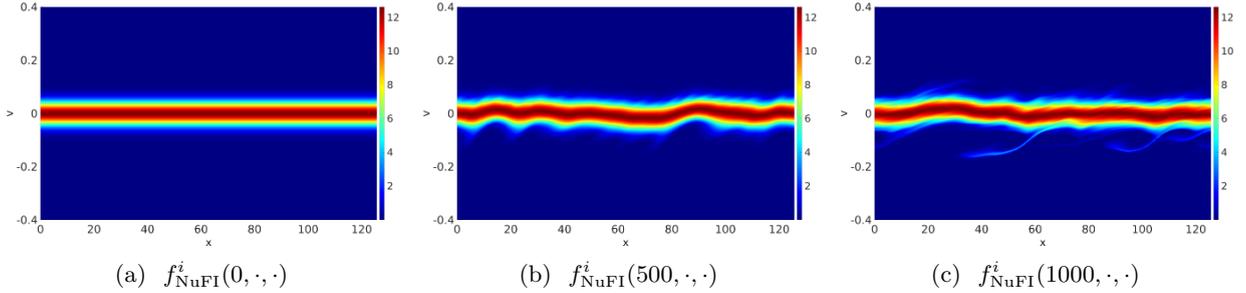

Figure 1: Distribution function $f_i$ of the ions at several stages in the simulation (computed using NuFI) shown for the *ion-acoustic turbulence* with a weak initial (quasi-random) perturbation, see Section 3.2.

### 2.6 Handling of velocity-support expansion

The initial data $f_0$ has to decay (quickly) for $|v| \to \infty$, therefore one can effectively say that $f_0$ has a bounded velocity support, i.e., is contained in $\Omega \times [v_{\min}, v_{\max}]$ up to a certain tolerance. While in many standard-benchmarks like *weak linear Landau damping* or the *two stream instability* for the single-species Vlasov–Poisson system the velocity support essentially remains the same even for large times, this is no longer true when going to two-species (or general multi-species) Vlasov–Poisson systems, where the velocity support can change drastically throughout the simulation period as can be seen in Figure 1.[8],[19]

Particle methods account for this support change by design as the particle represent the distribution function and therefore adapt by themselves. For NuFI, however, we have to adjust the velocity support *algorithmically*. Note, as the actual velocity support can vary strongly between different spatial positions $v_{\min}$ and $v_{\max}$ should be chosen depending on $x$.

A simple line-search at each $x$ to find $v_{\min}$ and $v_{\max}$ would be both very expensive in terms of the number of evaluations of $f(t, x, \cdot)$ needed and will tend to be inaccurate as one might miss fine filaments outgoing from the main distribution, the like of which can be seen in Figure 1c. Instead we suggest using that due to

$$\frac{\mathrm{d}}{\mathrm{d}s}\hat{v}(s) = -E(s, \hat{x}(s)), \tag{13}$$

from Equation (7), the velocity support can only change at most by

$$\tilde{E} = \max_{s \in [t, t+\Delta t]} |E(s, \hat{x}(s))| \tag{14}$$

during one time-step. Therefore if $f(t + \Delta t, x, v_{\max})$ is larger than a given tolerance one can set

$$v_{\max}^{\text{new}} = v_{\max} + \beta |E(t, x)|, \tag{15}$$

where $\beta \geq 1$ is introduced as safety-factor to account for numerical errors as well as potential underestimation due to $|E(t, x)|$ only being a rough estimate to $\tilde{E}$. Analogously one would set

$$v_{\min}^{\text{new}} = v_{\min} - \beta |E(t, x)|. \tag{16}$$

From Figure 1b to Figure 1c one can observe the 'folding' of outgoing filaments back towards the main distribution. Such 'holes' might be left unaccounted for when only using Equation (16) and Equation (15). Therefore one should additionally check for large gradients along $x$ in $\min(\mathrm{supp}(f(t, x, \cdot)))$ as well as $\max(\mathrm{supp}(f(t, x, \cdot)))$ and locally smooth them out if needed.

# 3 Numerical Experiments

In the following we will present results computed by NuFI in some benchmarks modelling both ions and electrons through a Vlasov equation with periodic boundary conditions. As a comparison for the quality of results we use an implementation of the Particle-in-Fourier (PIF) approach, which we chose due the auto-adaptive character of the underlying particle method and the reasonably good robustness with respect to noise of PIF in contrast to a classical PIC approach.[10]

This work's main focus lies on robustness and accuracy of NuFI in the multi-species setting rather than computational performance, therefore we restrict ourselves to the $d = 1$ case, where high resolutions can be achieved even using moderate hardware.

To investigate the suitability of NuFI for simulation of multi-species kinetic turbulences, which occur in e.g. *laser driven shock waves*, we, as a first step, want to consider the reproduction of *ion acoustic waves*. To this end we first look at a periodic setting and trigger ion acoustic waves by two different types of perturbations. The setup is taken from the work of Arber and Vann, where it was suggested as a benchmark for kinetic solvers.[19] The initial data is

$$f_0^e(x,v) = \frac{1}{\sqrt{2\pi}} p(x) \exp\left(-\frac{1}{2}(v - U_e)^2\right) \tag{17}$$

for electrons and

$$f_0^i(x,v) = \sqrt{\frac{M_r}{2\pi}} \exp\left(-\frac{M_r}{2} v^2\right) \tag{18}$$

for ions. Here $U_e = -2$ is the bulk velocity of the electrons, the mass ration between electrons and ions is set to $M_r = 1000$ and $p(x)$ is the perturbation for the electron distribution function.

Note that when we plot distribution functions computed by PIF one might encounter overshoots, i.e., regions where $f > \|f_0\|_{L^\infty}$, which is unphysical. Thus to avoid this, in a sense, plotting artefact, we set all plotted values of the distribution function, which are larger than allowed to the largest $\|f_0\|_{L^\infty}$.

## 3.1 Strong pertubation by a single mode

First we look at a perturbation by a single strong mode, similar to a setup which might be produced by a laser impulse. To this end we set

$$p(x) = \alpha \cos(kx) \tag{19}$$

with $\alpha = 0.5$, $k = 0.5$ and $x \in [0, L]$ with $L = 4\pi$.

As for NuFI the there is no intrinsic restriction on the time-step size (like a CFL condition for grid-based approaches) the time-step could be chosen as large as $\Delta t = 1/4$ while still remaining sufficiently accurate. To compute the integral for $\rho$ we employ the schemes presented in Section 2.5 and Section 2.6, where the maximum depth of refinement is set to $\theta_i = 3$ for the ions and to $\theta_e = 5$ for the electrons, while the minimal resolution is set to $N_v^e = N_v^i = 32$ for both. As the ion dynamics set in at later stage and tend to be less complicated during the simulated time period it was sufficient to use the lower depth to integrate over the ion distribution.

In Figure 2 the evolution of the electric energy over time is shown for both NuFI and PIF. While in the long run both methods seem to overall capture the right level of electric energy after the initial increase when choosing a sufficient resolution, the low-resolution PIF tends to somewhat under-predict the electric energy. Additionally in Figure 2a one can observe that PIF struggles to capture the right timings during the initial phase for lower resolutions. This is in good agreement with results previously observed in single-species benchmarks:[25] In the initial phase the electrons dominate the overall dynamics, resulting in what essentially is *non-linear Landau damping*. Intrinsic noise in a particle-based scheme complicates capturing the right timings, while NuFI reproduces these even with low resolutions, which we attribute to the structure preserving nature of NuFI.

This can be also observed when considering the convergence of NuFI and PIF respectively for $E$ and $\rho$ at fixed points in time: In Figure 3c and Figure 3d we display $\rho_e$ and in Figure 3e and Figure 3f $\rho_i$ for $t = 100$ and 300 computed using NuFI and PIF with a range of different resolutions. When $f^e$ becomes filamented one can see this in $\rho^e$, however, PIF seems to be less robust towards the filamentation resulting in more noise in $\rho^e$. To a lesser extend one can also observe this effect in $\rho_i$. For NuFI the overall behaviour is captured correctly even with low resolution and the resulting $\rho_e$ and $\rho_i$ computed by NuFI seem smoother than for PIF. This then also reflects in the reproduction quality of $E$, see Figure 3a and Figure 3b.

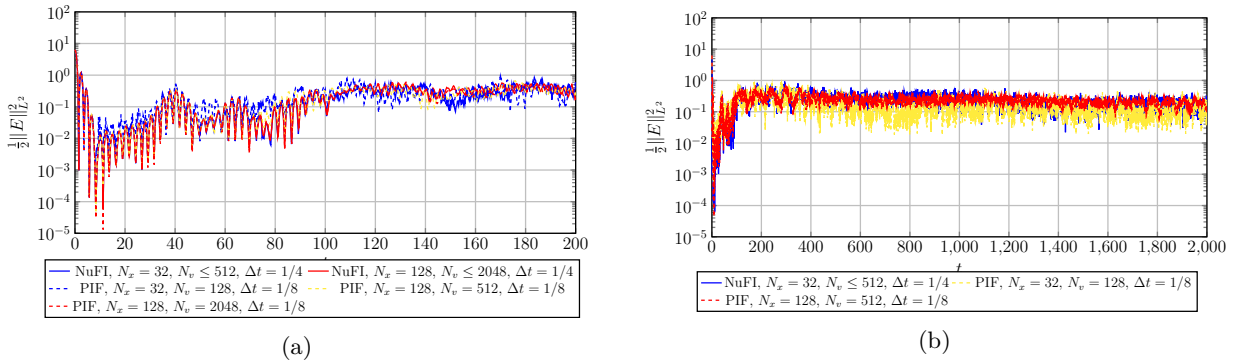

Figure 2: Electric energy for ion acoustic waves induced by strong initial perturbation compared between NuFI and PIF. Left is the electric energy displayed for the initial phase (until $t = 200$) and on the right is the electric energy over a longer period (until $t = 2000$).

In Figure 4c we display $f_e$ at $t = 2000$: fine structures are neither smeared nor completely overridden by noise and thus can still be clearly made out even for late times. This is not the case for PIF, see Figure 4f, where one can only qualitatively make out the shape of the distribution, while most details are lost to particle noise. Furthermore in Figure 4 the location of the vortex does not agree between NuFI and PIF.

In Figure 5 we make the comparison between $f^i$ computed by NuFI and PIF and observe a similar tendency as for $f^e$ in Figure 4, where while both methods are able to capture the dynamics qualitatively, the results of PIF are less accurate due to particle noise. Additionally for late times, after $t = 1000$, artefacts seem to appear in $f^i$ computed using PIF, see Figure 5e and Figure 5f.

## 3.2 Weak pertubation by background noise

Next we consider a perturbation by a superposition of several modes with less strength than in Section 3.1. This is similar to a quasi-random perturbation in the initial plasma state, however, we fix the modes and strength for reproducibility of the benchmark as suggested by Arber and Vann.[19] Set $p$ to

$$p(x) = \alpha \big( \sin(x) + \sin(0.5x) + \sin(0.1x) + \sin(0.15x) + \sin(0.2x) \\ + \cos(0.25x) + \cos(0.3x) + \cos(0.35x) \big) \tag{20}$$

with $\alpha = 0.01$ and $x \in [0, L]$ with $L = 40\pi$.

As the initial perturbation is significantly smaller in this benchmark inducing dynamics in the ions takes longer here, thus it also takes the ions longer to start effecting the electrons. The perturbation consists of a number of modes, which in turn leads to a rather 'chaotic' dynamic with strong filamentation in $f_e$. For PIF the filamentation manifests as numerical noise in the simulation, leading to strong deviations in computing $E$ even for early times, while NuFI is less prone to numerical noise, which can be observed in Figure 6b. Note that due to the high amount of excited modes in the solution one requires higher resolutions in $x$ to accurately capture the dynamics for longer times than in Section 3.1.

In Figure 6a, one can also observe that PIF struggles more than NuFI to capture the dynamics in the electric energy for comparable resolutions in the initial phase (until $t \approx 200$) of the simulation. However, when choosing 'insufficient' resolution in either $x$ or $v$ for NuFI it tends to over predict the electric energy for late times, thereby potentially introducing unphysical dynamics. We think that this might be caused by incorrectly predicting the velocity support and thereby missing parts of the distribution function when integrating. Even if initially severely over predicting the electric energy, PIF seems to allows relax back to approximately the right level of electric energy for large $t$. This effect needs to be further investigated.

In Figure 6c and Figure 6d we compare the electron distributions computed using NuFI and PIF respectively. In phase space both use $N_x = 128$ and for NuFI we allowed up to $128 \leq N_v^e \leq 2048$ for electrons and $32 \leq N_v^i \leq 128$ for ions, while for PIF we set $N_v^e = N_v^i = 1024$. The time step was chosen $\Delta t = 1/4$ for NuFI and $\Delta t = 1/16$ for PIF. Even though the resolutions were comparable in phase space and PIF was allowed to use smaller time-steps, the electron distribution function produced by PIF was very noisy practically not allowing to observe any structural details, which is not the case for the solution computed via NuFI where one can still clearly make out the vortices as well as some structure in the bulk of the electron distribution.

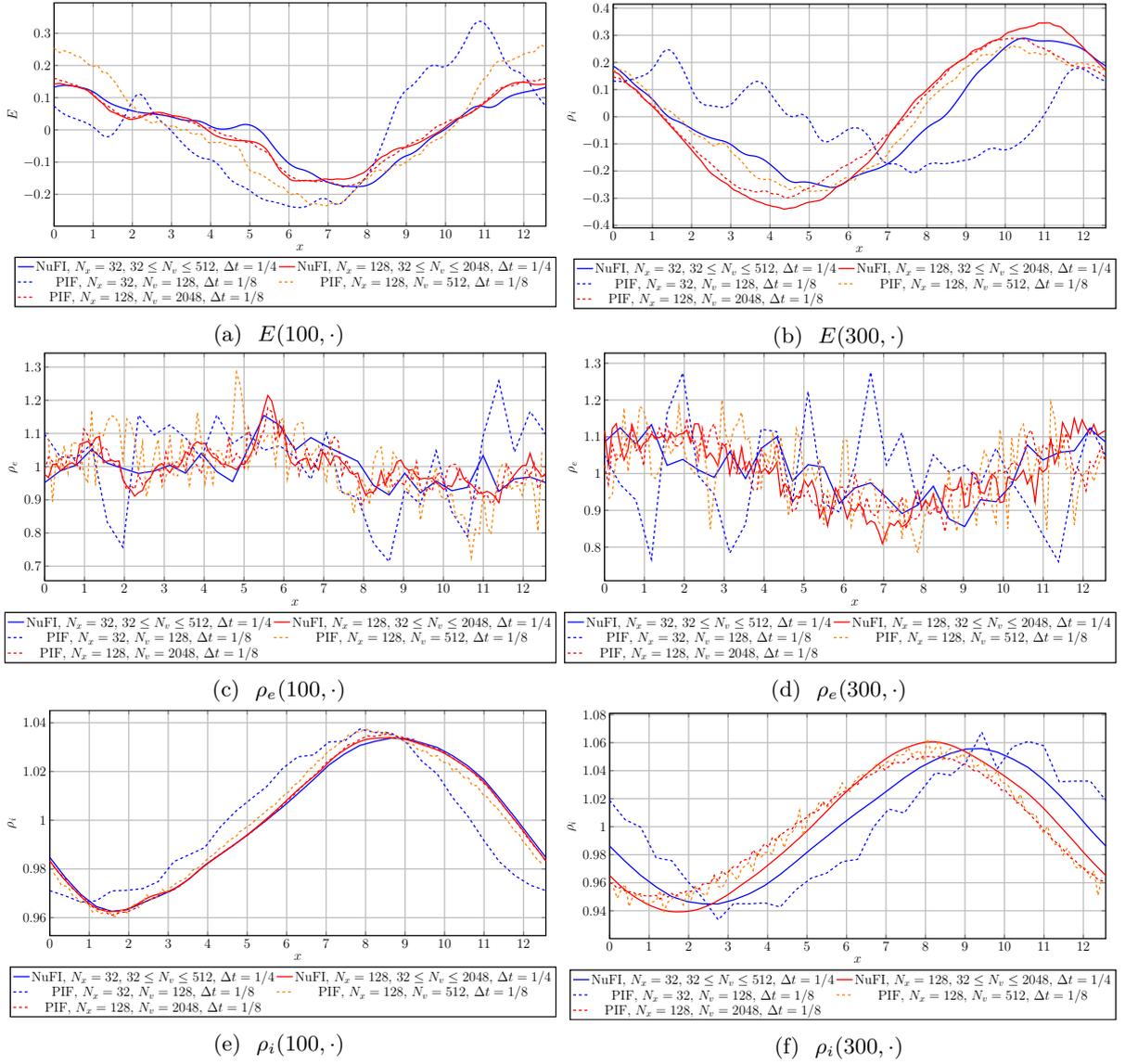

Figure 3: The figure shows $E$ (top), $\rho_e$ (middle) and $\rho_i$ (bottom) for the times $t = 100$ and $300$ for an ion acoustic wave induced by a single, strong perturbation. The results of low and high resolution NuFI (continuous line) and PIF (dotted line) simulations are compared.

## 4 Conclusion

In this work we have shown that NuFI can be used as a high fidelity simulation tool for multi-species plasma simulations in the electro-static, collisionless limit. It has been shown that when implementing adaptative integration and automatic tracking of the velocity support one can indeed run long-time stable simulations. This enables the user to profit from the noise-free character of grid-based approaches while also maintaining the adaptivity and conservation properties of particle-based approaches, resulting in a structure-preserving and accurate scheme for the Vlasov–Poisson system. In particular this leads to NuFI requiring lower resolutions and potentially enabling larger time steps while maintaining a good solution quality compared to other solvers.

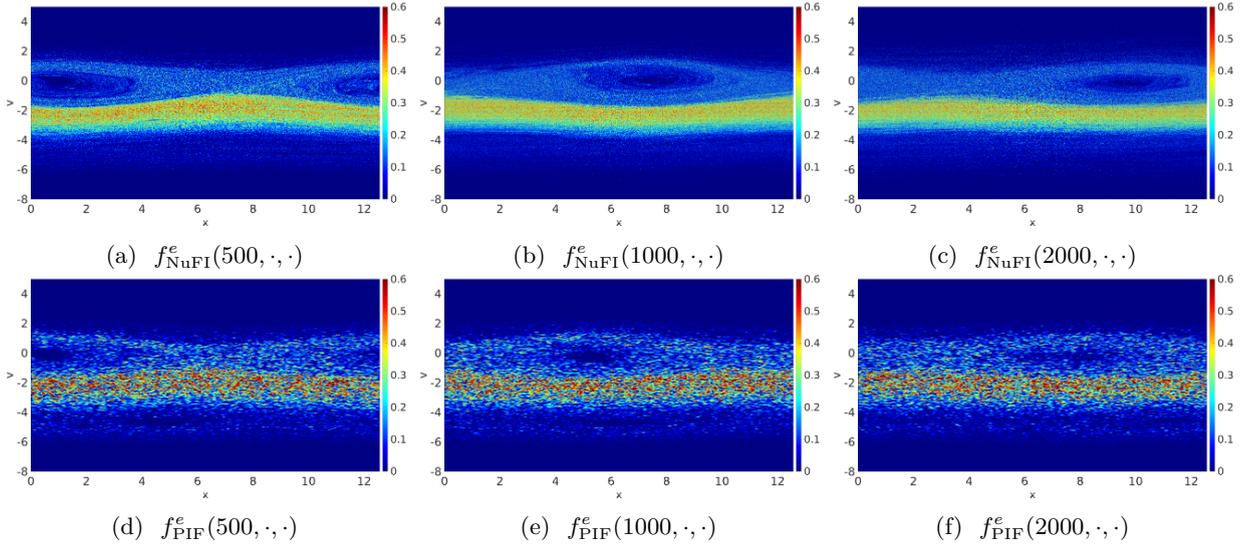

Figure 4: Distribution function $f^e$ of the electrons at times $t = 500, 1000$ and $2000$ for an ion acoustic wave induced by a single, strong perturbation. The upper 3 figures display the respective $f^e$ computed using (adaptive) NuFI with $N_x = 32$, $32 \leq N_v \leq 512$ and $\Delta t = 1/4$. The lower 3 figures display the respective $f^e$ computed using PIF with $N_x = 128$, $N_v = 512$ and $\Delta t = 1/8$.

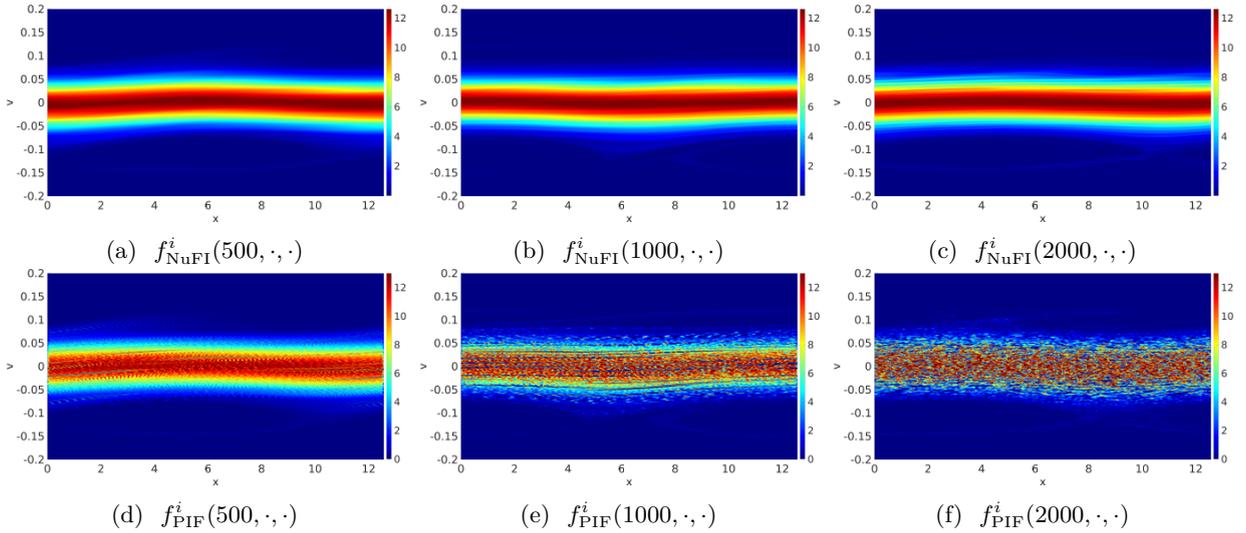

Figure 5: Distribution function $f^i$ of the ions at times $t = 500, 1000$ and $2000$ for an ion acoustic wave induced by a single, strong perturbation. The upper 3 figures display the respective $f^i$ computed using (adaptive) NuFI with $N_x = 32$, $32 \leq N_v \leq 128$ and $\Delta t = 1/4$. The lower 3 figures display the respective $f^i$ computed using PIF with $N_x = 128$, $N_v = 512$ and $\Delta t = 1/8$.

While it has been shown in previous works that NuFI benefits from its low memory consumption and high degree of parallelism allowing for almost optimal scaling,[22] for very long simulations this approach becomes again slower than e.g. PIF due to its quadratic scaling in the number of time-steps. Therefore we think that at the moment NuFI is best suited to simulate critical phases in plasma dynamics like transition regimes or formation of shocks where high precision is essential, after which one can switch back to models with less computational complexity. Additionally one can also use NuFI as 'cheap but accurate alternative' to storing the dynamics, i.e., instead of storing expensive snapshots of the distribution functions over time it suffices to only store the electric potentials over time.

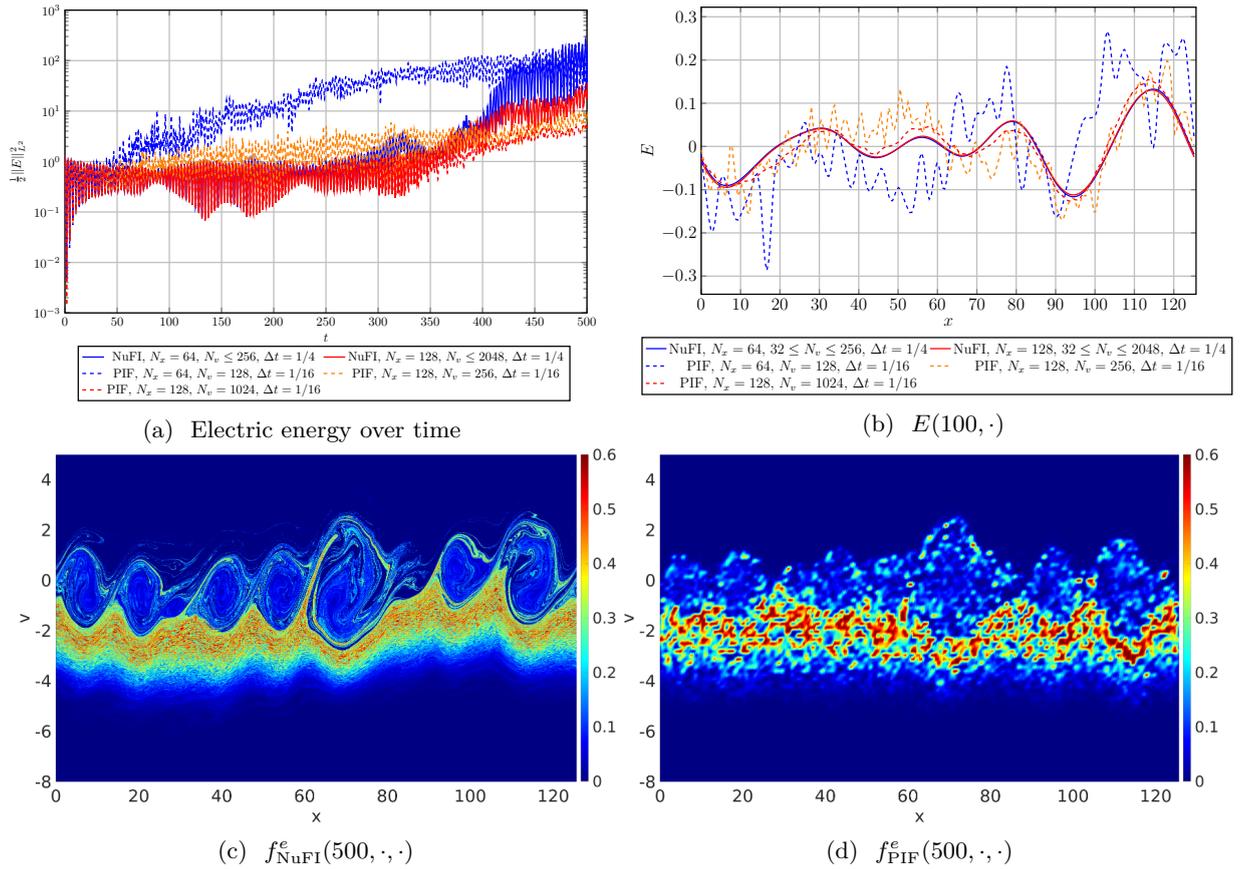

Figure 6: Results of simulating the ion acoustic benchmark provided by Arber and Vann.[19] Top left figure shows the evolution of electric energy over time. The top right figure compares the electric field for time $t = 100$ computed by NuFI and PIF respectively. The bottom figures compare the electron distribution computed by NuFI and PIF with comparable resolution in phase space and smaller time-steps for PIF.

In this work we focussed purely on solution quality, however, to be feasible for practical simulations it is also important to consider performance in terms of time-to-solution. While previous work suggested that for short to medium-long simulations NuFI is at least on par with PIF in terms of computation time to reach a required accuracy,[25] the question remains whether this also remains true for higher dimensions and more complicated setups. As the memory complexity of NuFI remains independent of the number of considered particle species, we think that NuFI might scale better than PIF in a multi-species setting. This should, however, be investigated in future research.

## Acknowledgments

Paul Wilhelm is recipient of a scholarship from the German national high performance computing organisation (NHR) funded by the Federal Ministry of Education and Research as well as the state governments. This work was partially funded by the Deutsche Forschungsgemeinschaft (DFG, German Research Foundation) – 333849990/GRK2379 (IRTG Hierarchical and Hybrid Approaches in Modern Inverse Problems).

# Disclaimer

This is the version of the article before peer review or editing, as submitted by the authors to the *IOP Special Issue on High Performance Supercomputing (HPC) in Fusion Research 2023.* IOP Publishing Ltd is not responsible for any errors or omissions in this version of the manuscript or any version derived from it.